# Identification of urinary biomarkers of food intake for onion by untargeted LC-MS metabolomics

Qian Gao[1], G. Gürdeniz[1], Giulia Praticò[1], Camilla T. Damsgaard[1], Eduvigis Roldán-Marín[2], M. Pilar Cano[3], Concepción Sánchez-Moreno[2], Lars O. Dragsted[1*].

[1] *Department of Nutrition, Exercise and Sports, University of Copenhagen, Copenhagen, Denmark*

[2] *Institute of Food Science, Technology and Nutrition (ICTAN), Spanish National Research Council (CSIC), Madrid, Spain*

[3] *Department of Biotechnology and Food Microbiology, Institute of Food Science Research (CIAL) (CSIC-UAM), Madrid, Spain*

**Abbreviations:** BFI, biomarker of food intake; PLSDA, partial least squares discriminant analysis; PLS, partial least squares regression; NSC, nearest shrunken centroid; ROC, receiver operator characteristic; SVM, support vector machines; PCA, principal component analysis; AUROC, the area under the receiver operator characteristic curve; VIP, variable importance in projection; ACSOs, S-alk(en)yl-L-cysteine sulfoxides; CPMA, N-acetyl-S-(2-carboxypropyl)-cysteine; ECHS1D, enoyl-CoA hydratase deficiency; NAGS, N-acetylglutamate synthase; UGTs, UDP-glucuronosyl transferases.

**Abstract**

**Scope:** Biomarkers of food intake (BFIs) are useful tools for objective assessment of food intake and compliance. The aim of this study was to discover and identify urinary BFIs for onion.

**Methods and results:** In a randomized controlled cross-over trial, 6 overweight participants (age 24-62 years) consumed meals with 20 g/d onion powder or no onion for 2 weeks. Untargeted UPLC-qTOF-MS metabolic profiling analysis was performed on urine samples and the profiles were analysed by multilevel-PLSDA, modified PLS, and nearest shrunken centroid to select features associated with onion intake. Eight biomarkers were tentatively identified; six of them originated from S-substituted cysteine derivatives such as isoalliin and propiin, which are considered the most specific for onion intake. Most of the biomarkers were completely excreted within 24 hours and no accumulation was observed during 2 weeks indicating their ability to reflect only recent intake of onions. Receiver-operator curves were made to evaluate the performance of individual biomarkers for predicting onion intake. The area under the curve values for these biomarkers ranged from 0.81 to 1.

**Conclusion:** Promising biomarkers of recent onion intake have been identified in human urine. Further studies with complex diets are needed to validate the robustness of these biomarkers.

# 1. Introduction

Onion (*Allium cepa*) is one of the major horticultural crops and is cultivated around the world [1]. It is a common food ingredient used in almost all cultures and has been applied as a medicinal food since ancient times [2]. Onion consumption has been associated with decreased risk of cardiovascular disease and may be anti-hypertensive, hypocholesterolemic, and hypolipidemic [3, 4]. However, only few studies have been conducted in humans and the results are controversial.

Several studies suggest that the proposed health-promoting effects of onion are due to its high content of organosulfur compounds, flavonoids and dietary fiber [4, 5]. The major organosulfur compounds in onion are γ-glutamylcysteine and various S-substituted alkyl cysteine derivatives, which are the precursors of the compounds producing the characteristic odor and flavor of onion [6]. Onion also contains numerous flavonoids and among which quercetin derivatives are the most prevalent ones across all onion cultivars followed by derivatives of kaempferol and isorhamnetin [7]. Onion is also featured with high contents of soluble fiber such as inulin and fructo-oligosaccharides [8]. These three groups of compounds are considered as the constituents most likely to be responsible for the proposed health effects, however such effects need to be further confirmed in human studies.

In order to establish the association and further cause and effect relationship between onion and health, it is important to have an accurate measurement of onion consumption in free-living individuals. Biomarkers of food intake are a robust tool to assess dietary intake in observational studies and to monitor compliance in intervention studies [9, 10]. The discovery of new biomarkers of food intake will improve the assessment of dietary intake and provide a strong tool for better understanding the diet-health associations and the effect of food components. Up to date, there are few reported good biomarkers for assessment of onion intake. LC-MS-based untargeted metabolomics approaches have been shown as a powerful tool to identify new biomarkers in both observational and intervention studies [11, 12]. This metabolic profiling approach is able to detect thousands of metabolites in a single biological sample within a run generating large amounts of multivariate data. The metabolic profile consists of a large number of features, which are characterized by their intensity, retention time and mass to charge ratio (m/z), and may be sensitively analysed with a relatively small sample size. However, the large number of variables compared to samples increases the risk of overfitting the data and may result in misleading conclusions. It is not practical to increase the sample size to match the number of features because it would be costly and time-consuming. Therefore, a powerful classification method followed by a feature selection method is crucial to extract important information precisely even with few observations.

The aim of the present study was to identify biomarkers of food intake for onion using an LC-MS based untargeted metabolomic approach on samples from a randomized, controlled dietary intervention study. Three classification-feature selection methods - multilevel PLSDA, modified PLS and nearest shrunken centroid (NSC) – were adopted to provide a more precise selection of important features. The multiple sampling time points of urine after intervention allowed us to observe the time-response relationship of the candidate biomarkers.

## 2. Experimental Section

### 2.1. Study design

A randomized double-blinded cross-over study was conducted with two periods of 2 weeks and a 24-26 days wash-out between periods. At the beginning of each period, participants were given a test meal with onion powder or control to assess acute effects of onion. In the rest of each period, subjects partially substituted their lunch and dinner meals with 2 daily intervention meals containing onion powder (20 g/d) or control (no onion). The inclusion criteria for participants were: healthy persons aged 18-70 years, non-smoking, BMI>25 kg/m$^2$, exercise <10 hours/week, alcohol intake <21 units/week. Exclusion criteria were: cancer within the last 6 months, systemic and immunological diseases, infections, inflammatory or other chronic diseases of the gut, serious psychiatric diseases, use of lipid lowering medication, non-steroid anti-inflammatory or acetylsalicylic acid containing drugs affecting oxidative capacity, concomitant participation in other research trials, recent weight loss >2 kg, restrictive diets, dietary supplements (except multivitamins) blood donation, pregnancy, or breastfeeding.

To minimize the influence of sulphur compounds and flavonoids from the background diet, subjects were instructed to follow a restriction diet for 7-9 d before and during each intervention period, and received a recipe booklet to help them follow the diet. The restriction diet excluded all *Allium* vegetables (including garlic, spring onions, leek etc.), cabbage, radish, asparagus, rocket, watercress and fruits such as apple, pear, banana, berries, grapes and stone fruits, jam, honey, chocolate, sweets containing fruit, drinks such as tea, drinking chocolate and fruit juice (pineapple juice and 1 glass/d of orange juice were excepted), most alcoholic drinks, balsamic and red wine vinegar, ready-made salad dressings, mustard, ketchup, and spice mixtures containing onion or garlic. Small amounts of other spices were accepted. Compliance with the restriction diet was ascertained at each visit. Apart from the dietary restrictions and intervention meals, subjects were asked to maintain their habitual diet, lifestyle and physical activities. The study was conducted according to the guidelines laid down in the Declaration of Helsinki and all procedures involving human subjects were approved by the Committees on Biomedical Research Ethics of the Capital Region of Denmark (no. H-D-2008-058).

**2.2. Subjects**

Ten overweight but apparently healthy subjects (5 men and 5 women) were recruited through advertisements and six (3 men and 3 women) of them completed the study. They had a median age of 43 (range 24-62) years, a mean BMI of 31.5 ± 3.3 kg/m$^2$ and their waist-to-hip ratios ranged 0.78-0.96 (women) and 1.06-1.09 (men). Mean blood pressures were 129 ± 13/ 82 ± 8 mmHg and plasma total cholesterol 4.9 ± 1.4 mmol/l. Two participants took a multivitamin tablet daily at enrolment and were instructed to continue during the study. One participant took a gastric acid inhibitor (40 mg/d Inexium) to reduce gastric reflux daily during the study.

**2.3. Onion powder**

The freeze-dried onion powder was produced at the Department of Plant Food Science and Technology at *Institute of Food Science, Technology and Nutrition (ICTAN), Spanish National Research Council (CSIC)*, Madrid, Spain. Raw onions (*Allium cepa* L. var. *cepa*, 'Recas') were peeled and freeze-dried to obtain a fine ground onion powder of which 20 g corresponded to 100 g fresh onion. The quercetin content was 3.4 ± 0.5 mg/g, equivalent to a daily dose of about 136 mg in the onion intervention period.

**2.4. Meals**

The test meal at the beginning of each intervention period consisted of a double serving of potato soup with 20 g onion powder or control served with 120 g white bread, and 300 mL water. The meal contained 2622 kJ energy, 12.5 g protein, 24.5 g fat and 91.2 g carbohydrate. Control meals contained 8.5 g sucrose and 2 g soy protein instead of onion powder.

The intervention meals in the rest of each intervention period were provided frozen to the subjects and to be eaten at home. Four different meals were produced in one batch: tomato soup, potato soup, minced meat pot and meat loaf. Subjects were asked to eat one soup and one meat dish per day, corresponding to 20 g/d freeze-dried onion powder (about 100 g/d fresh onion) or no onion. The subjects were instructed to heat the meals and not to share them with others or leave anything uneaten. Suggestions for accompaniments such as pasta, rice, bread and non-restricted vegetables (cucumber, carrot, tomato etc.) were given.

**2.5. Sample collection**

24 h pooled urine samples were collected during 2 days before the intervention period. At the beginning of each intervention period, urine samples were collected before and after each test meal covering the following time periods: 1) before the test meal; 2) 0-2 h after the test meal; 3) 2-4 h after the test meal; 4) 4-24 h after the test meal. In addition, samples from the four time points in the day with test meal were mixed to create 24 h pooled samples. Finally, 24 h pooled urine samples were collected for the last 2 days of the intervention period. The overview of the time points for sample collections are shown in Figure S1.

**2.6. UPLC-QTOF-MS analysis**

Urine samples were thawed in a refrigerator or on ice and centrifuged at 3000 g for 2 min at 4 °C. Then 150 μL of each supernatant was randomized and placed in the wells of 96-well plates where samples from the same subject were kept in the same plate. Samples were analysed by an ultra-performance liquid chromatography (UPLC) system coupled to quadruple time-of- flight (Premier QTOF) mass spectrometer (Waters Corporation, Manchester, UK), as described previously [13]. The mobile phase was 0.1 % formic acid in water as solvent A and 0.1 % formic acid in 70:30 (v/v) ACN:MeOH as solvent B. HSS T3 $C_{18}$ column (2.1 × 100 mm$^2$, 1.8 μm) was used and a gradient of mobile phase solvent and flow rate was applied for a run time of 7 min. Electrospray ionization (ESI) was used for the analysis and the capillary probe voltage was 3.2 and 2.8 kV for positive mode and negative mode, respectively. In full scan mode, the selected m/z range was from 50 to 1000 Da. Ion source and desolvation gas temperatures were 120 and 400 °C, respectively.

### 2.7. Data preprocessing

The raw data were converted to NetCDF files using DataBridge (Waters, Manchester, UK). NetCDF files were imported into MZmine 2.19 for data preprocessing following the steps below: mass detection, chromatogram builder, chromatogram deconvolution (local minimum search), isotopic peaks grouper, join aligner (peak alignment), duplicate peak filter, peak list rows filter and peak finder (gap filling). Each resulting feature represents an ion with a specific retention time and m/z; the signal intensities of the features were defined by the peak heights [14].

MZmine preprocessed data were imported into MATLAB R2015b ver. 8.6.0.267246 (Mathworks Inc., MA, USA). The following steps were applied separately on datasets from positive and negative ionization mode. Feature reduction was applied with the following steps: First, the features with the retention time before 0.3 min and after 6 min were removed. Second, a '70 % rule' was applied to remove the features except the ones which had peak intensity higher than 20 and were detected in more than 70 % of the samples in at least one of the subgroup (divided according to meals and time points). Third, the features with a coefficient of variation (CV) higher than 0.7 were removed [15].

Then probabilistic quotient normalization [16] was performed to correct the urine concentration variations among samples. The mean intensity of each metabolite within the same plate was set to its overall mean of all samples to remove inter-batch variation. In order to facilitate identification, features detected within 0.01s and with correlation coefficients higher than 0.7 were grouped and considered to come from the same compound [15].

### 2.8. Data Analysis

All data analysis was performed in MATLAB R2015b ver. 8.6.0.267246 (Mathworks Inc., MA, USA) using codes developed based on MLPLSDA [17], N-way toolbox [18] and NSC [19].

Multilevel PLSDA, modified PLS and NSC were applied on the preprocessed and auto-scaled data and the features selected by all three methods were extracted for further evaluation and identification.

#### 2.8.1. Multilevel PLSDA

Preprocessed data were subjected to multilevel PLSDA [17]. The between subjects variation and within subjects variation were separated. Principal component analysis (PCA) was performed on whole data and on the data including within subjects variation to evaluate the effect of separating the within subjects variation. Then PLSDA was applied on the within subject variation to find discriminating features between the two treatments. A single cross validation scheme was adopted to develop the PLSDA model. Training sets were created by setting aside all the samples of two subjects and in total, 15 training sets were developed by running all the possible of combinations of two subjects. Variable selection was performed based on variable importance in projection (VIP) scores and the selectivity ratio (the cut-off value for both was 1) until the cross-validation classification error did not increase anymore. Permutation test was applied to assess the classification performance by comparing misclassifications, AUROC and $Q^2$ between 10000 random class assignments and original classification.

**2.8.2. Modified PLS**

A modified PLS regression with group × time-response as independent Y was applied (Gao et al., in preparation). In brief, a single cross validation was performed on the preprocessed data to determine the optimal number of latent variables for a PLS model. Balanced bootstrapping was performed to resample 200 bootstrap datasets and a PLS model with the optimal number of latent variables was built on each bootstrap subset. A VIP score was obtained for each variable in each model. The mean VIP and standard deviation for each variable were calculated across all the models. If the lower-bound of the standard deviation error bar was above 1, the variable was selected.

**2.8.3. NSC**

NSC is a classification method based on nearest centroid classification, where the distances between the class centroid and the overall centroid of features are step-wise reduced (shrunk) by a threshold to obtain an optimal number of features resulting the highest true class probability or lowest classification error [20]. NSC with a double cross validation scheme was applied on preprocessed data, as described previously [19]. In the inner loop, class centroids and overall centroids of the training set were calculated. The discriminant and probability score were calculated for each test sample and each shrinkage value. In the outer loop, the probability models obtained from inner loops were applied on the test set and the discriminate and true class probability scores were calculated. The true class probability for each sample was summed up and the optimal shrinkage was determined based on the maximum true class probability of the test set. Because the optimal shrinkages obtained from different outer loops were not always the same, curves of the true class probability and errors vs shrinkage were plotted and the shrinkage at the maximum of the median curve (having minimal error) was used as optimal shrinkage. This optimal shrinkage was applied to the whole data set and the corresponding features were selected.

**2.8.4 Evaluation of biomarkers**

Receiver operator characteristic (ROC) curves with associated confidence intervals are considered as the most statistically valid approach for the assessment of biomarker performance [21]. Balanced bootstrap was used to resample 200 bootstrap subsets from preprocessed data. ROC curve for each individual selected feature was built on each subset using support vector machines (SVM) as classifier. Mean area under the ROC curve (AUC) and their confidence intervals were calculated across 200 bootstrap subsets.

**2.9. Identification**

Full-scan MS spectra of selected features were inspected and parent ions were identified based on adducts and fragments. The selected features were compared to an in-house metabolite database and MS/MS fragmentation analysis (10, 20, and 30 eV) was applied as described previously on unknown compounds to provide more information regarding their structures [22]. The resulting spectra were compared to literature and searched in the following online databases: HMDB (http://www.hmdb.ca), FooDB (http://foodb.ca/), Chemspider (http://www.chemspider.com), Metlin (https://metlin.scripps.edu), Chembank (http://chembank.broadinstitute.org/welcome. htm) and MetFrag (https://msbi.ipb-halle.de/MetFrag/). The level of identification has been assigned as described by Sumner et al. [23].

## 3. Results

In total, there were 84 urine samples (6 subjects × (four time points + three pools)) for analysis. After data preprocessing and feature reduction, 3209 (+) and 3521 (-) features remained for further analysis.

### 3.1. Statistical analysis

#### 3.1.1. Multilevel PLSDA

The total variation was split into offset, between-subject variation and within-subject variation. PCA was applied to visualize the total variation (before split) and within-subject variation (after split) in data obtained from samples collected at 2 h after the test meal (Figure 1). For both positive and negative mode, onion and control samples were separated better in within-subject variation than in the total variation indicating the intervention effect was mainly reflected in within-subject variation and that subjects may react differently to onion. PLSDA was subsequently applied on the within-subject variation and variable selection was performed based on VIP scores and selectivity ratios. In total 322 and 354 features were selected for the positive and negative mode, respectively. A permutation test was adopted to validate the multilevel PLSDA model. Compared to the misclassifications and AUROC of 10000 permutations with random class labelling, the misclassifications and AUROC calculated from the original datasets were significantly different ($P < 0.05$) (Figure 2). Results of PCA scores plots and permutation tests on data generated from samples obtained at other time points are shown in Figures S2-S5.

### 3.1.2. Modified PLS

PLS with group×time as response Y was applied on the data. 200 times bootstrap resampling were performed and VIP scores were calculated for each feature. The bootstrap-VIP scores for the first 200 features with highest mean VIPs are shown in Figure 3. The features in red were considered relevant because the lower-bound of the standard deviation of the VIP score was higher than 1. Not all the features with high mean VIPs were selected due to their large variation between bootstrap subsets. 170 and 199 features were selected for positive and negative mode, respectively.

### 3.1.3. NSC

The sum of true probabilities and classification error as a function of shrinkage for data obtained at different time points are shown in Figure 4 and Figure S6-S10. The optimal shrinkages observed in two plots are consistent and spot samples (2h and 4h) showed lower variability than pooled samples. The NSC approach selected 109 and 121 features for positive mode and negative mode, respectively.

### 3.1.4. Feature selection

The overview of the feature selection by the three statistical analysis methods is shown in Table S1. Multilevel PLSDA selected the largest number of features with a large proportion of unique features. NSC was the most conservative method and resulted in the smallest number of selected features, most of which were also selected by other methods. In total, 96 and 69 features were selected by all three methods for positive and negative mode, respectively. These features were considered to be associated with onion consumption and were selected for further identification.

### 3.2. Identification of biomarkers of food intake for onion

The features associated with onion intake by all three methods were grouped by correlation analysis at their respective retention times to form a total of 85 groups of related features, i.e. belonging to maximally 85 different distinct metabolites. Eleven of the feature groups identified as putative biomarkers of food intake for onion were structurally elucidated belonging to eight distinct metabolites (Table 1) and their structures are shown in Figure 5. The information on unknown metabolites is given in Table S2.

Isoalliin and S-(2-carboxypropyl)-cysteine (metabolite 1 and 2) were identified as biomarkers originating directly from onion itself. Both of them peaked around 2 h after onion consumption and showed similar excretion profiles (Figure S11). Six metabolites were identified as biomarkers originating from metabolism. The identification was based on their MS/MS spectra as shown in Figure S13-S15. Their short term excretion profiles are provided in Figure S11. Most of these metabolites (3, 4, 6, 7) peaked between 0-2 h while metabolite 5 and 8 peaked in urine collected between 2 and 4 h. The putative precursors of these biomarkers and their biotransformation processes are shown in Figure 5. The putative precursors of biomarkers 3-7 are S-substituted cysteine derivatives while biomarker 8 was originating from isorhamnetin. During the two week intervention period, no significant changes were observed in the concentrations of these eight metabolites (Figure S12).

## 4. Discussion

In this study, eight metabolites were found and tentatively identified as biomarkers of food intake for onion using an LC-MS-based untargeted metabolomics approach. These metabolites originated from either onion or human metabolism and showed increased concentration in urine after onion consumption in the intervention study.

*Allium* vegetables contain various S-alk(en)yl-L-cysteine sulfoxides (ACSOs), such as S-(prop-2-enyl)cysteine sulfoxide (alliin), S-(prop-1-enyl)cysteine sulfoxide (isoalliin), S-(methyl)cysteine sulfoxide (methiin) and S-(propyl)cysteine sulfoxide (propiin), etc. [24]. They are flavor precursors and precursors of organosulfur compounds [1]. The occurrence of these compounds vary across *Allium* species and are highly specific to the species [25]. Isoalliin is the dominating sulfur containing cysteine sulfoxides in onion while alliin is dominating in garlic. It has been reported that molar ratios of methiin/alliin/isoalliin for onion was 16/0/84 while for garlic was approximately 17/78/5 [26]. The structures of alliin and isoalliin are isomers and the only difference is the position of the C=C double bond. Although the analytical method in the current study was not able to discriminate alliin and isoalliin, we assume it was isoalliin that has been detected in the current study based on its prevalent distribution in onion. It is worth noting that, when *Allium* vegetables are cut or chewed, alliinase is released from the vacuole and ACSOs are hydrolyzed to some extent to their corresponding alk(en)yl sulfenic acids in the damaged area [27]. Therefore, these compounds are unstable. They are also sensitive to cooking processes, leading to low concentrations in normal *Allium*-containing meals. Isoalliin was observed to have a relatively high concentration in the current study, possibly due to the fact that the onion was freeze-dried and served as onion powder. The freeze-drying process might deactivate the alliinase and prevent the degradation of isoalliin. The change of the concentration of isoalliin is also complicated by the existence of its corresponding γ-glutamyl peptides. It has been suggested that γ-glutamyl- S-(prop-1-enyl)cysteine is converted to isoalliin by γ-glutamyl transpeptidase and oxidase during storage [28] and steaming [29] of onion. The application of isoalliin as biomarker of food intake is affected by the food processing method. Therefore, further studies with cooked and raw whole onion must be performed to investigate its potential use as a biomarker of onion intake.

S-(2-carboxypropyl)-cysteine is an S-substituted cysteine derivative, which is recognized as biosynthetic precursors of a range of important sulfur containing compounds in onion. It is biosynthesized from cysteine and methacrylic acid, and provides alk(en)yl groups (except methyl) to some ACSOs and S-(2-Carboxypropyl)-glutathione, among others [1]. Its presence in baseline and control samples indicates that there are other sources of this compound besides onion (and most fruit and vegetable) consumption. It has been proposed to be a metabolite of methacrylyl-CoA, which is a reactive intermediate in the valine catabolic pathway [30]. To our knowledge, S-(2-carboxypropyl)-cysteine has not been reported previously as a metabolite after onion intake.

N-acetyl-S-(prop-1-enyl)-cysteine sulfoxide, N-acetyl-S-(propyl)-cysteine sulfoxide and N-acetyl-S-(2-carboxypropyl)-cysteine were tentatively identified based on their MS/MS spectra and comparison with the structures of compounds originating from onion. Based on their structures, we hypothesize that they are metabolites of isoalliin, propiin and S-(2-carboxypropyl)-cysteine, respectively. No known pathway was found to be responsible for this transformation but it was assumed e it is an analogy to the N-acetyl conjugation happening in humans after consumption of garlic, where S-allyl cysteine is transformed into N-acetyl-S-allylcysteine with N-acetyl transferase(s) [31, 32]. N-acetyl-S-(prop-1-enyl)cysteine sulfoxide has been identified in 24 h pooled urine after onion consumption using NMR in a previous intervention study [33] and the structure identified in the present study is in accordance with their finding. Propiin is found to be abundant in chive while it is present in much small amounts in onion and only having trace contents in garlic [34]. Based on its variable distribution in *Allium* vegetables, its metabolite, N-acetyl-(propyl)-cysteine sulfoxide, has the potential to become a biomarker which discriminates different *Allium* species. To the best of our knowledge, this is the first time that this compound has been reported as a metabolite after onion intake. Its extremely low concentration in baseline samples and in the control group indicate that there is no major endogenous source of it. N-acetyl-S-(2-carboxypropyl)-cysteine (CPMA) has been reported by Jandke and Spiteller as a metabolite of onion and garlic [31]. They suggested γ-glutamyl-S-(2-carboxypropyl)-cysteinylglycine might be the precursor of CPMA since mercapturic acids often come from their corresponding glutathione conjugates. However, similar to its precursor, CPMA might also come from methacrylyl-CoA, the intermediate of valine catabolism already mentioned [35]. It has been suggested as a biomarker for short-chain enoyl-CoA hydratase deficiency (ECHS1D) diagnosis because the level of it in urine increased even in the mildest cases of ECHS1D [36]. However, the level of CPMA is low in the healthy population as observed in the baseline samples and in samples from the control treatment in the current study, indicating that it could be a useful biomarker of onion intake in the healthy population.

2-Acetyliminopropanoic acid is a small molecule and is part of the structure of isoalliin. It also present as a fragment in the MS/MS spectra of N-acetyl-S-(prop-1-enyl)cysteine sulfoxide and showed similar retention time as it. Therefore, we suspect that this molecule highly likely came from the fragmentation process. Nevertheless, it is a featured structure for the metabolites of S-substituted cysteine derivatives and it has not been observed after the control treatment or in previous studies. As a structure that is easily obtained from the fragmentation of other metabolites associated with onion intake, we believe it could potentially serve as a biomarker of onion intake.

N-Acetylglutamic acid is known as an endogenous metabolite which comes from the reaction between glutamic acid and acetyl-CoA by N-acetylglutamate synthase (NAGS) [37]. Glutamic acid is common in plant- and animal-based food such as chicken, egg and soy beans, etc. [38]. The concentration of N-acetylglutamic acid was stable in the control group where no restriction was applied on the glutamic acid source in the diet. This indicates that dietary glutamic acid might not be a source of N-acetylglutamic acid in urine. The reason that the concentration of it was higher in urine of subjects who consumed onion might be because of the prevalent existence of γ-glutamyl groups in onion. Further studies are needed to confirm this finding.

Isorhamnetin 4'-O-glucuronide is the conjugation metabolite of isorhamnetin-4'-O-glucoside, which is particularly abundant in onion and shallot and can be efficiently absorbed into the small intestine after consumption [39]. Like the metabolism of the analogous quercetin derivative, during the absorption the glucoside is hydrolyzed and isorhamnetin released for glucuronidation by UDP-glucuronosyl transferases (UGTs). Isorhamnetin 4'-O-glucuronide has previously been detected in plasma shortly after the ingestion of onion [39]. Due to the extensive distribution of quercetin and in many plant based food such as apple, grape, and tea, etc., isorhamnetin 4'-O-glucuronide, which is also a quercetin metabolite in humans lacks specificity as a biomarker of intake for any specific food, but can be used as compliance biomarker of onion intake in highly controlled intervention studies such as the current study.

The time-response relationship within 24 h after a single exposure of the eight metabolites was also observed in this study. All metabolites had short half-lives and peaked around 2 h after onion consumption except for N-acetyl-S-(2-carboxypropyl)-cysteine and isorhamnetin 4'-O-glucuronide, which both peaked in urine collected between 2-4 h. The optimal time window for measurement of these biomarkers would therefore be 2-4 h after intervention. The 24 h pooled urine samples would also be a feasible choice. No change of concentrations of the eight metabolites was observed during the two-week intervention period indicating stable metabolism and lack of accumulation of these metabolites in the human body after repeated onion intake.

In conclusion, eight biomarkers in urine were identified to be associated with onion intake and their time-responses after a single exposure were observed. None of them changed upon repeated exposures for 14 days. Two metabolites excreted unmetabolised (isoalliin, S-(2-carboxypropyl)-cysteine) and four metabolites were products of their metabolism (N-acetyl-S-(prop-1-enyl)-cysteine sulfoxide, N-acetyl-S-(propyl)-cysteine sulfoxide, N-acetyl-S-(2-carboxypropyl)-cysteine) and 2-acetyliminopropanoic acid were considered as biomarkers that are specific for onion intake. N-acetylglutamic acid and isorhamnetin 4'-O-glucuronide can be used to monitor compliance in intervention studies under certain conditions. Further validation studies in free-living subjects are needed to confirm the robustness of these biomarkers and studies with common cooking of onion would be needed to validate and investigate for additional biomarkers.

**Figure legends**

**Figure 1.** PCA scores plot on total variation (top) and within-subject variation (bottom) from data generated in positive mode (left) and negative mode (right) on samples obtained at 2 h after test meal.

**Figure 2.** Number of misclassifications, area under the ROC curve and Q2 calculated on permuted data (histogram) and original data (red dot) generated in positive mode (left) and negative mode (right) on samples obtained at 2 h after test meal.

**Figure 3.** Bootstrap-VIP scores for first 200 variables with highest mean VIPs in data generated in positive mode (top) and negative mode (bottom), respectively. Red and black represent the variables which are selected or not-selected as relevant features in the method. The horizontal blue line corresponds to VIP=1.

**Figure 4.** Probability plot (upper left) and error plot (lower left) of the NSC model and their corresponding descriptive statistics (right) based on 100 repetitions of the double cross validation scheme on data generated in positive mode on samples obtained at 2 h after the test meal. Red, mean curve; blue, median curve; green, curves for mean ± standard deviation.

**Figure 5.** Structures of compounds in onion and metabolites detected in urine after onion consumption (1-8). 1 Isoalliin; 2 S-(2-carboxypropyl)cysteine; 3 N-acetyl-S-(prop-1-enyl)cysteine sulfoxide; 4 N-acetyl-S-(2-carboxypropyl)cysteine; 5 N-acetyl-S-(propyl)cysteine sulfoxide; 6 2-Acetyliminopropanoic acid; 7 N-Acetylglutamic acid; 8 Isorhamnetin 4'-O-glucuronide.

**Figure 6.** Excretion profiles of 3: N-acetyl-S-(prop-1-enyl)cysteine sulfoxide, 4: N-acetyl-S-(propyl)cysteine sulfoxide, and 6: 2-acetyliminopropanoic acid in urine.

**Author contributions**

LOD was responsible for study design, blood and urine analyses. CTD coordinated the study. ER-M was responsible for production of the onion powder. CSM and MPC were responsible for study design and onion powder production. QG was responsible for the MS/MS analysis of the samples, the statistical analysis and interpretation of the data. GG, GP and LOD were involved in the interpretation of the data. QG drafted the manuscript and revised it after receiving the comments from other co-authors. All authors read and commented on the manuscript.

**Acknowledgements**

This work was supported by the Danish Ministry of Food, Agriculture and Fisheries (NuBI, 3304-FVFP-060696-01) and the Spanish Ministry of Science and Innovation (Consolider-Ingenio Programme 2010, FUN-C-FOOD, CSD2007-00063; AGL2010-15910/ALI and INIA RTA2015-00044-C02-02). The work was funded by a grant from the China Scholarship Council (201506350127) to QG.

**Conflict of interest**

The authors declare no conflict of interest.

**Table 1.** Identified urine metabolites originated from onion and metabolism.

| No | Mode | RT | m/z | Molecular formular | AUC (95 % CI) | Metabolite identification[a] | Source |
|---|---|---|---|---|---|---|---|
| 1 | POS | 0.63 | 178.059 | C6H11NO3S | 1 (1, 1) | Isoalliin[II] | Onion |
| 2 | POS | 0.83 | 208.065 | C7H13NO4S | 1 (1, 1) | S-(2-carboxypropyl)cysteine[II] | Onion |
| 3 | POS | 1.31 | 220.065 | C8H13NO4S | 1 (1, 1) | N-acetyl-S-(prop-1-enyl)cysteine sulfoxide[II] | Metabolism |
|   | NEG | 1.32 | 218.047 |  | 1 (1, 1) |  |  |
| 4 | POS | 1.57 | 222.079 | C8H15NO4S | 1 (1, 1) | N-acetyl-S-(propyl)cysteine sulfoxide[III] | Metabolism |
|   | NEG | 1.57 | 220.063 |  | 1 (1, 1) |  |  |
| 5 | POS | 2.71 | 250.076 | C9H15NO5S | 1 (1, 1) | N-acetyl-S-(2-carboxypropyl)cysteine[II] | Metabolism |
|   | NEG | 2.71 | 248.060 |  | 1 (1, 1) |  |  |
| 6 | POS | 1.31 | 130.056 | C5H7NO3 | 1 (1, 1) | 2-acetyliminopropanoic acid[III] | Metabolism |
| 7 | POS | 0.81 | 190.068 | C7H11NO5 | 0.81 (0.75, 0.88) | N-Acetylglutamic acid[I] | Metabolism |
| 8 | NEG | 3.55 | 491.085 | C22H20O13 | 0.82 (0.76, 0.88) | Isorhamnetin 4'-O-glucuronide[I] | Metabolism |

[a] Identification level according to Sumner et al [23].

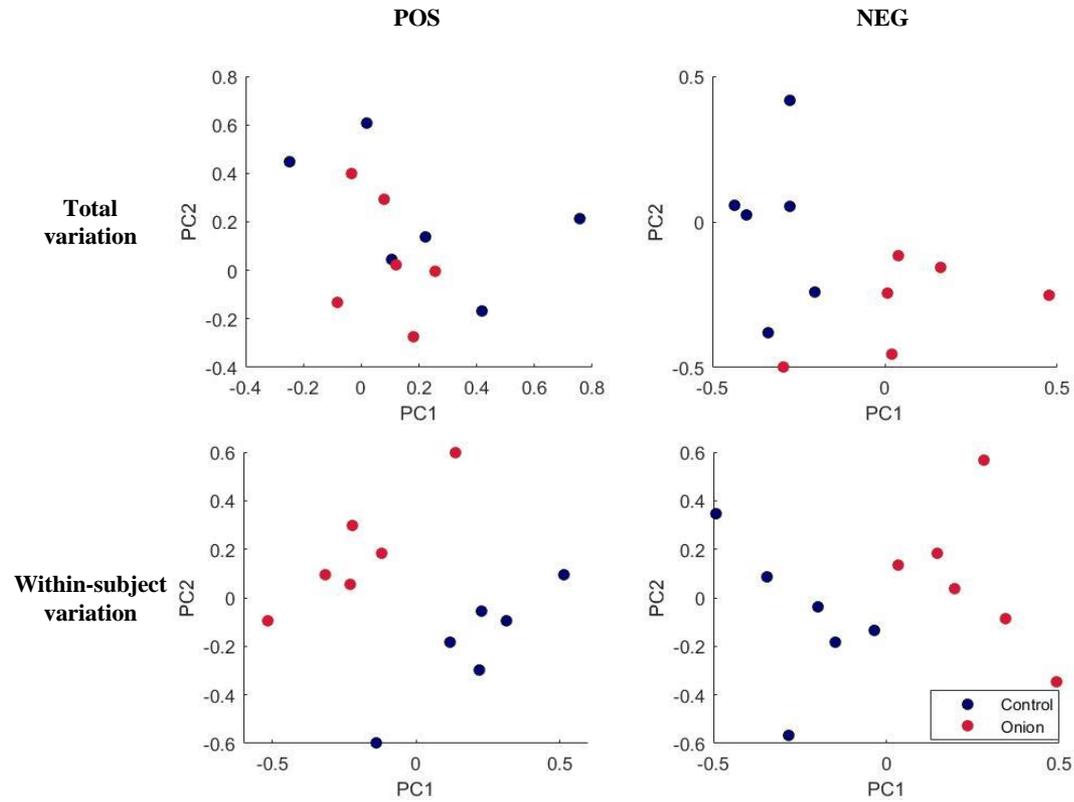

**Figure 1.** PCA scores plot on total variation (top) and within-subject variation (bottom) from data generated in positive mode (left) and negative mode (right) on samples obtained at 2 h after test meal.

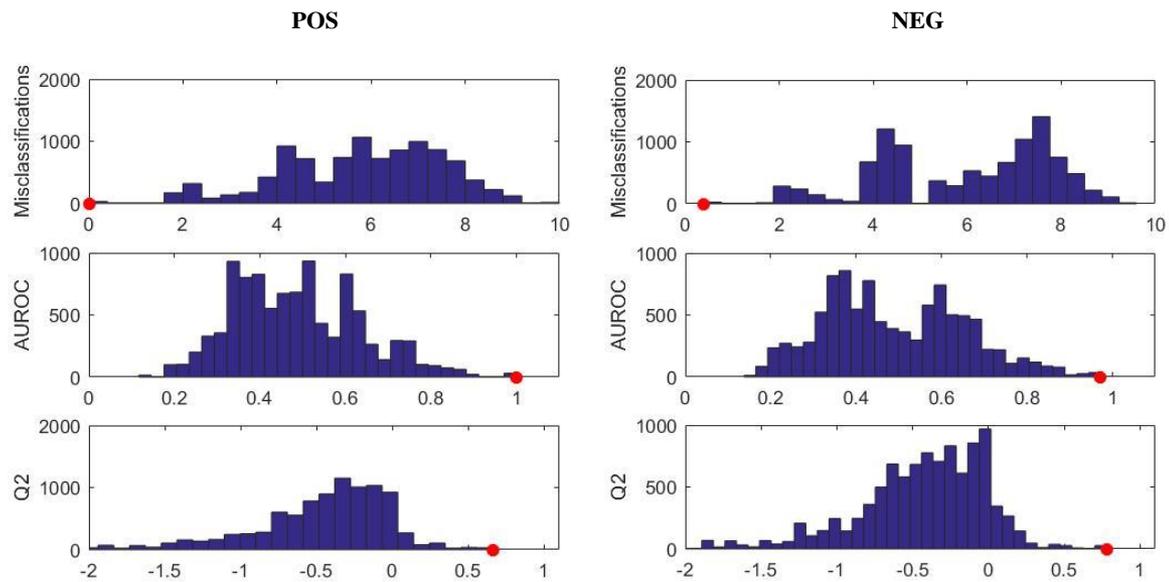

**Figure 2.** Number of misclassifications, area under the ROC curve and $Q^2$ calculated on permuted data (histogram) and original data (red dot) generated in positive mode (left) and negative mode (right) on samples obtained at 2 h after test meal.

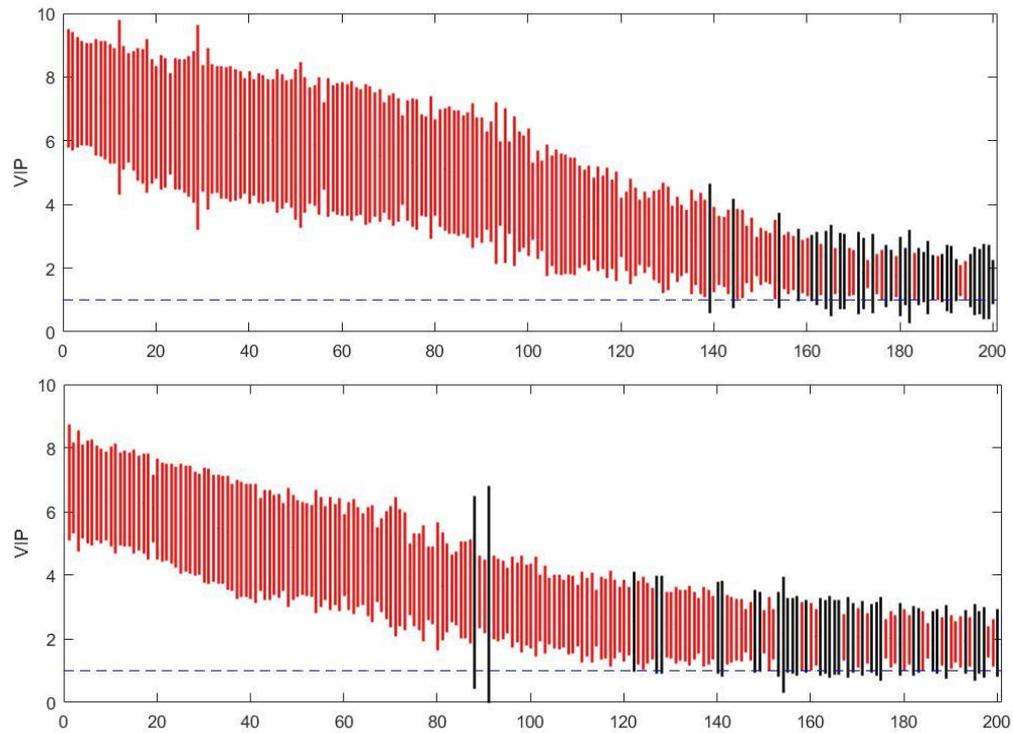

**Figure 3.** Bootstrap-VIP scores for first 200 variables with highest mean VIPs in data generated in positive mode (top) and negative mode (bottom), respectively. Red and black represent the variables which are selected or not-selected as relevant features in the method. The horizontal blue dash line corresponds to VIP=1.

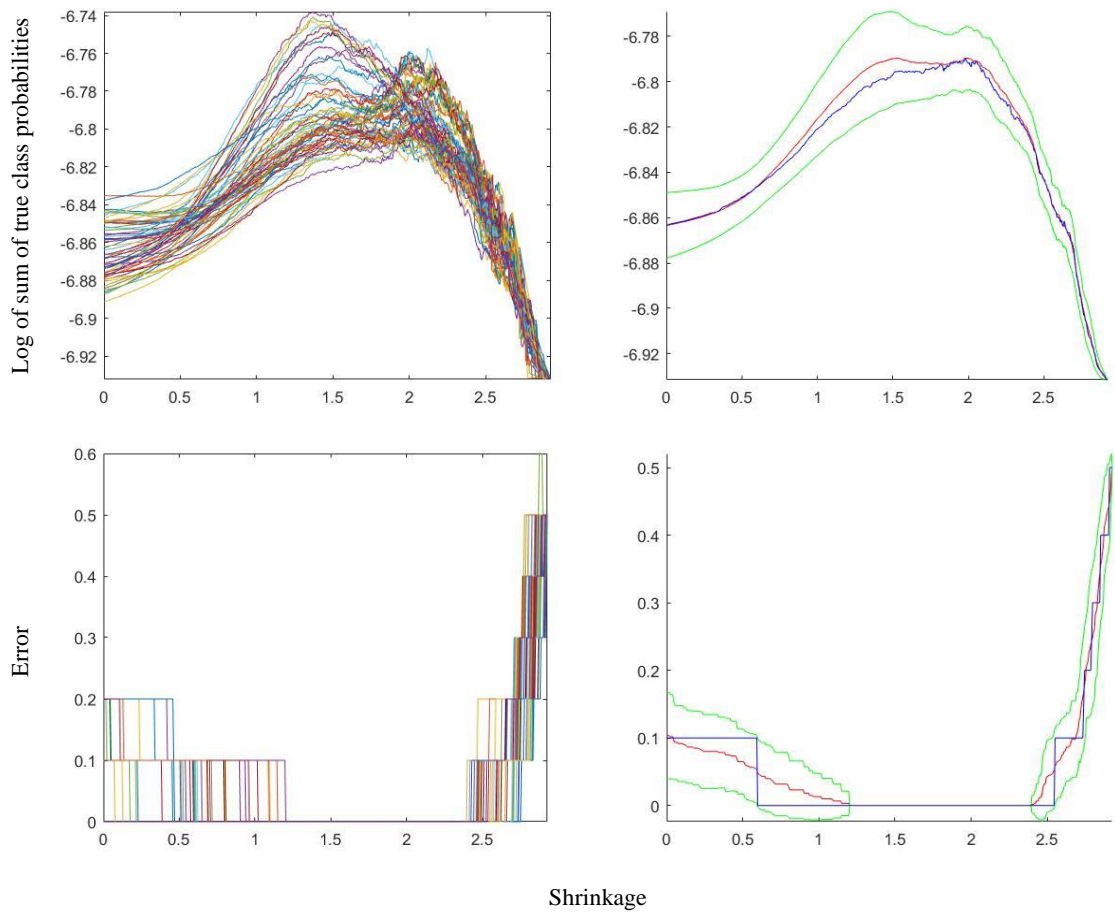

**Figure 4.** Probability plot (upper left) and error plot (lower left) of the NSC model and their corresponding descriptive statistics (right) based on 100 repetitions of the double cross validation scheme on data generated in positive mode on samples obtained at 2 h after the test meal. Red, mean curve; blue, median curve; green, curves for mean ± standard deviation.

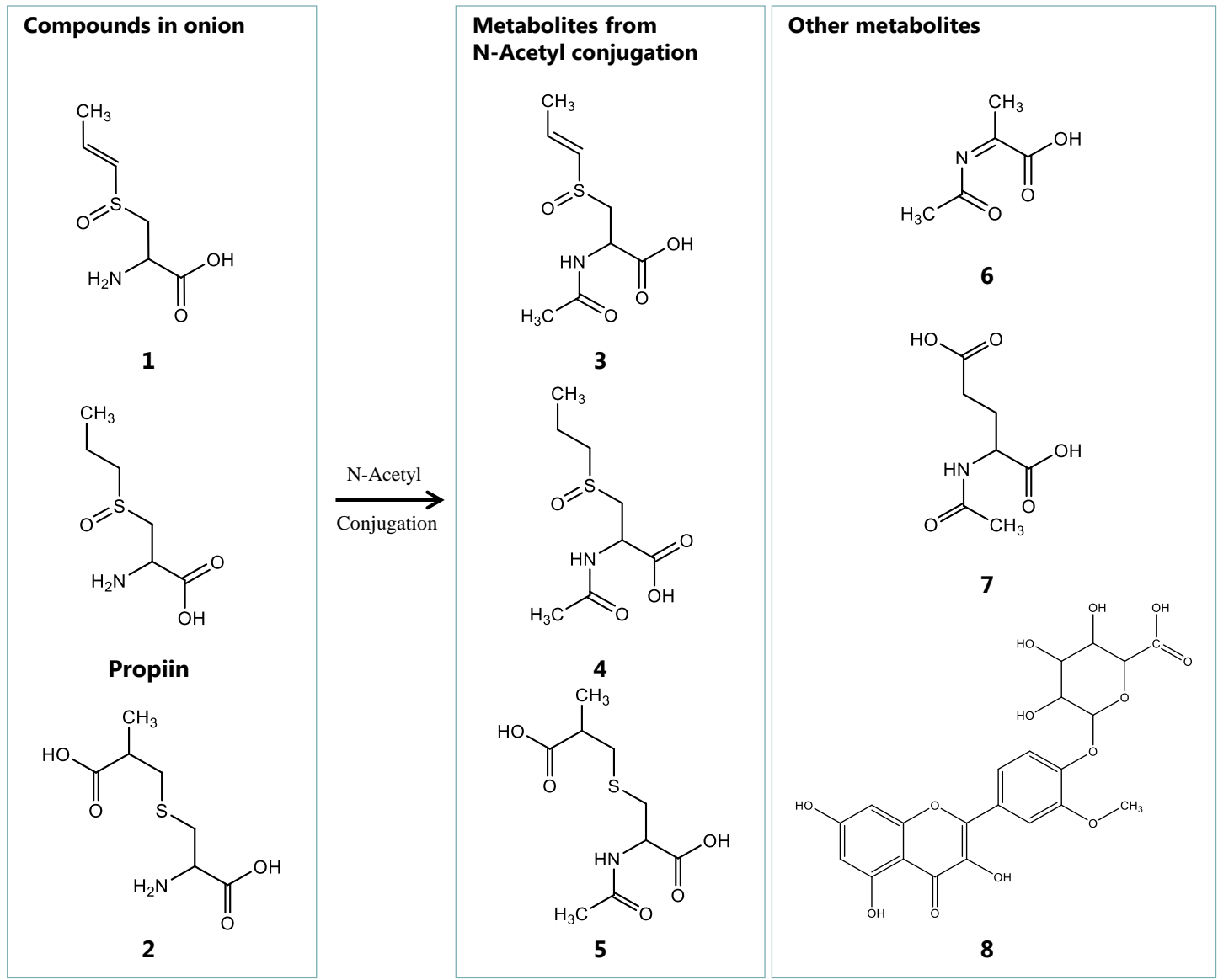

**Figure 5.** Structures of compounds in onion and metabolites detected in urine after onion consumption (1-8). 1 Isoalliin; 2 S-(2-carboxypropyl)cysteine; 3 N-acetyl-S-(prop-1-enyl)cysteine sulfoxide; 4 N-acetyl-S-(propyl)cysteine sulfoxide; 5 N-acetyl-S-(2-carboxypropyl)cysteine; 6 2-Acetyliminopropanoic acid; 7 N-Acetylglutamic acid; 8 Isorhamnetin 4'-O-glucuronide.

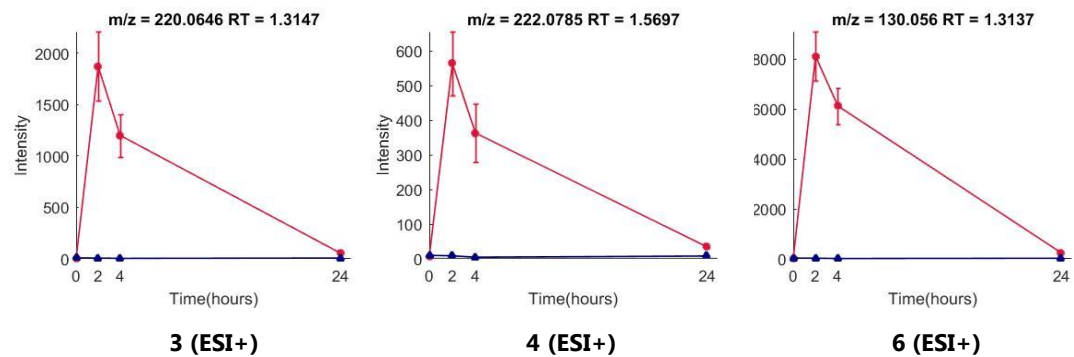

**Figure 6.** Excretion profiles of 3: N-acetyl-S-(prop-1-enyl)cysteine sulfoxide, 4: N-acetyl-S-(propyl)cysteine sulfoxide, and 6: 2-acetyliminopropanoic acid in urine.